\documentclass[12pt,a4paper]{article}

\usepackage{amsfonts} 
\usepackage{braket} 
\usepackage{caption} 
\usepackage{subcaption} 
\usepackage{amssymb} 
\usepackage{amsmath}
\usepackage{xcolor}
\usepackage{graphicx}
\usepackage[hyphens]{url}
\usepackage{doi}
\usepackage{hyperref}
\usepackage[top=25mm,bottom=25mm]{geometry}

\newtheorem{theorem}{Theorem}[section]
\newtheorem{remark}[theorem]{Remark}

\definecolor{matlabblue}{rgb}{0,0.447,0.741}
\definecolor{matlabred}{rgb}{0.85, 0.325, 0.098}
\definecolor{matlabyellow}{rgb}{0.929, 0.694, 0.125}
\definecolor{matlabpurple}{rgb}{0.494, 0.184, 0.556}
\definecolor{matlabgreen}{rgb}{0.466, 0.674, 0.188}

\begin{document}
	
	\title{How to coordinate vaccination and social distancing to mitigate SARS-CoV-2 outbreaks\thanks{This work was funded by the Federal Ministry of Education and Research (BMBF; grants 05M18EVA and 05M18SIA).}}

\author{%
	Sara Grundel%
	\thanks{Max Planck Institute for Dynamics of Complex Technical Systems, Magdeburg, Germany
		(\texttt{[grundel,ritschel]@mpi-magdeburg.mpg.de}).
	}
	\and
	Stefan Heyder%
	\thanks{Technische Universit{\"{a}}t Ilmenau, Ilmenau, Germany, Institute for Mathematics 
		(\texttt{[stefan.heyder,thomas.hotz,philipp.sauerteig,karl.worthmann]@tu-ilmenau.de}).
	}
	\and
	Thomas Hotz\footnotemark[3]
	\and
	Tobias K. S. Ritschel\footnotemark[2]
	\and
	Philipp Sauerteig\footnotemark[3]\;\,\footnote{Corresponding author.}
	\and
	Karl Worthmann\footnotemark[3]
}

\maketitle

\begin{abstract}
Most countries have started vaccinating people against COVID-19. %
However, due to limited production capacities and logistical challenges it will take months/years until herd immunity is achieved. %
Therefore, vaccination and social distancing have to be coordinated. %
In this paper, we provide some insight on this topic using optimization-based control on an age-differentiated compartmental model. %
For real-life decision making, we investigate the impact of the planning horizon on the optimal vaccination/social distancing strategy. %
We find that in order to reduce social distancing in the long run, without overburdening the healthcare system, it is essential to vaccinate the people with the highest contact rates first. %
That is also the case if the objective is to minimize fatalities provided that the social distancing measures are sufficiently strict. %
However, for short-term planning it is optimal to focus on the high-risk group. %
\end{abstract}

\section{Introduction}
In early 2020, the outbreak of the \emph{severe acute respiratory syndrome coronavirus 2} (SARS-CoV-2) was declared a pandemic by the World Health Organization~\cite{WHO20}. As the virus causes the respiratory illness \emph{coronavirus disease 2019} (COVID-19), many countries have been enforcing nonpharmaceutical countermeasures such as social distancing (also called contact restrictions) and travel restrictions~\cite{IMF20c, IMF20b}. %
Such countermeasures have a severe negative impact on national and international economies~\cite{OECD20} and the general quality of life (in particular, mental health). Consequently, significant effort has been made to develop and deploy vaccines. %
However, in some countries, the initial vaccine rollout has been slower than anticipated, new strains of the virus are emerging, and public skepticism towards COVID-19 vaccines prevails~\cite{ChaCLetal21, Leo21}. Furthermore, experts warn that nonpharmaceutical measures remain necessary~\cite{Hun21, Par20} even as the vaccines are being deployed. Therefore, there is still a need for identifying strategies for safely relaxing nonpharmaceutical measures. %

One of the most popular methodologies for modeling the spread of a virus are so-called \emph{compartmental} models consisting of difference or differential equations~\cite{Het00}. %
Optimal control of such models is an active research area~\cite{Beh00, BolBS17, WatNP20}, and optimal mitigation policies have been proposed for several diseases, including dengue fever~\cite{FiscChud19} and malaria~\cite{OlaOAetal20}. %
A variety of approaches has been proposed for studying the impact of nonpharmaceutical measures on the spread of SARS-CoV-2. They range from network-based over game-theoretical to data-driven methods~\cite{PareBeck20, MengLore20, BrauMind20}. %
Since the beginning of the pandemic, many researchers have proposed compartmental models for predicting the impact of countermeasures on the spread of SARS-CoV-2~\cite{GioBBetal20,WiljGane20}. 
Besides prediction, these models can be used to determine optimal mitigation policies based on solving \emph{optimal control problems} (OCPs).
For instance, many authors have proposed SARS-CoV-2 mitigation strategies based on optimal control of nonpharmaceutical countermeasures (in particular, social distancing)~\cite{CarCEetal20, KohSJetal20, TsaLSetal20, GrimMeng20}.
Nonpharmaceutical measures were used to attain a stable equilibrium with low case numbers in~\cite{ContDehn20}. %
In~\cite{BinCris20}, the authors propose an on-off (also called bang-bang) social distancing policy for mitigating a second wave. The subject of optimal vaccination has also been considered~\cite{Nei20, GraLL20}.
Early work on optimal vaccination involved age-uniform vaccination, and on-off policies were found to be optimal~\cite{LibLPetal20}. %
Later, Matrajt et al.~\cite{MatELetal20} presented optimal \emph{age-targeted} vaccination policies in the absence of social distancing. They either consider the vaccination to be completed instantaneously at the initial time or assume constant vaccination rates. When minimizing deaths, they observe that for low vaccine efficacy, it is optimal to vaccinate the elderly. For high vaccine efficacy, and when sufficiently many are vaccinated, it is optimal to vaccinate the younger age groups which account for the most transmissions.
Similarly, Hogan et al.~\cite{HogWWetal20} consider optimal age-targeted vaccination where the entire age group is assumed to be vaccinated at a constant rate over the course of one month.
Acuna-Zegarra et al.~\cite{AcuDBetal20} use optimal control to conclude that %
intense vaccination for a limited time period is optimal. %
Buckner et al.~\cite{BucCS20} optimize time-varying age-targeted vaccination. In particular, they account for essential workers, e.g. health care professionals, which are unable to significantly reduce their social interaction. They find that, depending on the objective function, either 1)~younger essential workers are prioritized in order to control the spread of SARS-CoV-2 or 2)~senior essential workers are prioritized to control the mortality.
Finally, Bertsimas et al.~\cite{BerIJetal20} consider age- and region-differentiated vaccination tailored for a number of states in the US. They model the social distancing as a predefined function of time which is fixed in advance.
However, to the best of our knowledge, simultaneous optimal control of vaccination and social distancing has not been considered previously.

Furthermore, most work on optimal SARS-CoV-2 mitigation involves \emph{open-loop} strategies where a single OCP is solved using a long-term horizon.
However, due to the significant uncertainty surrounding SARS-CoV-2 and COVID-19, open-loop strategies are not sufficient because over time, the model parameters may change and the state of the pandemic will deviate from that predicted by the model. These issues can be addressed by using \emph{model predictive control} (MPC)~\cite{GruP17, RawMD19}, see also~\cite{CoroGrue20} for continuous-time systems and~\cite{WortRebl14} for the relation between continuous- and discrete-time systems. In MPC, the current state of the system is measured or estimated, an OCP is solved over a finite prediction and control horizon, and the first part of the solution is implemented in practice before repeating the procedure for a shifted horizon. This is referred to as a closed-loop strategy. It is similar to real-life decision-making where plans, e.g. for contact and travel restrictions, are reevaluated at regular intervals. %
MPC is a well-established control methodology, which has been applied successfully in several fields~\cite{ForPHetal15, mayne2014model, QinB03}. For applications of MPC in, e.g. power electronics, we refer to~\cite{vazquez2014model, cortes2008predictive}, in robotics to~\cite{WortMehr16} and the references therein, and in energy to~\cite{BrauGrue16,JianSaue20}. %
Finally, a suitable choice of the prediction horizon is essential, see e.g.~\cite{Wort11, EsteWort21}. %

In this work, we address four key questions related to coordinating social distancing and vaccination: 1)~How important are the availability and the success rate of the vaccine? 2)~Who should be vaccinated first? 3)~How much can social distancing measures be relaxed once the vaccines are available? 4)~What is the minimal prediction horizon in the optimization step that recovers the qualitative features of the long-term policy?
In order to address these questions, we present a novel compartmental model, which extends a recently developed model~\cite{GrunHeyd20} to account for vaccination.
Based on this model, we present optimal simultaneous vaccination and social distancing policies involving optimal control and MPC. %
The model accounts for vaccination failure, different levels of symptom severity, and age-dependent characteristics of SARS-CoV-2 and COVID-19.
In our case study, we choose parameters tailored to the COVID-19 outbreak in Germany. However, we expect that the conclusions carry over to other developed countries.

We observe that it is optimal to first vaccinate the middle-aged group which spreads the disease the most. Subsequently, the elderly  (the high-risk group) are vaccinated. In this case, the contact restrictions can be lifted almost half a year earlier than without vaccination. They can be lifted even earlier by increasing the number of \emph{successful} vaccinations, which depends on both the number of available vaccines and their success rate. Conversely, the maximal occupancy of the intensive care units (ICUs), which is closely related to the number of fatalities, can be reduced by \emph{prolonging} the contact restrictions instead (without making them more strict).
These conclusions are based on a prediction horizon of 2~years in the optimization step. We demonstrate that the same conclusions hold if a prediction horizon of at least 8~weeks is used. If the planning horizon is too short, the elderly are vaccinated first. However, this is short-sighted and requires more strict contact restrictions than if the middle-age group is vaccinated first.

The remainder of this paper is structured as follows. We present the compartmental model in Section~\ref{sec:novel:seiphr:model} and describe the optimal control problem in Section~\ref{sec:optimal_control}. Section~\ref{sec:results} is dedicated to the case study, and the paper is concluded in Section~\ref{sec:conclusions}.

\section{A Compartmental Model with Social Distancing and Vaccination}\label{sec:novel:seiphr:model}
In this section, we extend the dynamical model tailored to \mbox{COVID-19} proposed in~\cite{GrunHeyd20} %
by incorporating vaccination as an additional control input. %
To this end, all compartments are split into two parts distinguishing vaccinated from non-vaccinated people. %

Different vaccines have different properties. %
For simplicity, we focus on active vaccination, i.e. the body is triggered to produce antibodies itself. %
As a consequence, this kind of vaccination yields immunity but only if the patient has not been infected at time of vaccination. %
Still, there might be patients whose bodies do not produce (a sufficient amount of) antibodies. %
We assume that everyone has the same probability of vaccination failure and that a second try would yield the same outcome. %
Therefore, we allow vaccination at most once per person. %
These considerations motivate the following assumptions.
\begin{itemize}
	\item[\rm{(A1)}] Everyone who is not known to have been infected can be vaccinated. Vaccination can only be successful for people who have not been infected.
	\item[\rm{(A2)}] No one is vaccinated twice.
	\item[\rm{(A3)}] Each vaccination (of a non-infected person) has the same probability of failure.
	\item[\rm{(A4)}] Successful vaccination yields immediate immunity (no future infection possible). %
\end{itemize}
In~\cite{GrunHeyd20}, we proposed a SEITPHR model consisting of 11~compartments which are divided into $n_\mathrm{g}$ age groups, $n_\mathrm{g} \in \mathbb{N}$. %
The compartments account for susceptible ($S$), exposed (or latent) ($E$), infectious ($I$), tested ($T$), hospitalized with and without requiring an intensive care unit (ICU) ($P$ and $H$), and detected and undetected removed people ($R^K$ and $R^U$). %
The term \emph{removed} captures all people who neither infect others nor need an ICU in the future, i.e. recovered, deceased, and quarantined people without severe infection. %
In order to account for symptom severity, we split~$I$ into $I^S$, $I^M$, and~$I^A$, where the superscripts $S$, $M$, and~$A$ indicate whether a person has a severe course of infection, i.e. he or she will go to an ICU at some point in time, a mild course, i.e. he or she will show symptoms and be put into quarantine and therefore will be removed, or an asymptomatic one, i.e. he or she will not be detected at all. %
Our model does not account for reinfection, i.e. recovered people cannot get infected again. %
Furthermore, we do not distinguish between people with mild symptoms and people with severe symptoms who refuse intensive care. %
In Appendix~\ref{app:hosp_min}, we consider optimal vaccination based on the minimization of the total number of fatalities. %
There, for simplicity, we assume a constant ratio between total number of people treated on an ICU and fatalities.

\begin{figure}[htbp!]
\begin{subfigure}{0.9\textwidth}
	\centering
	\includegraphics[width=\textwidth]{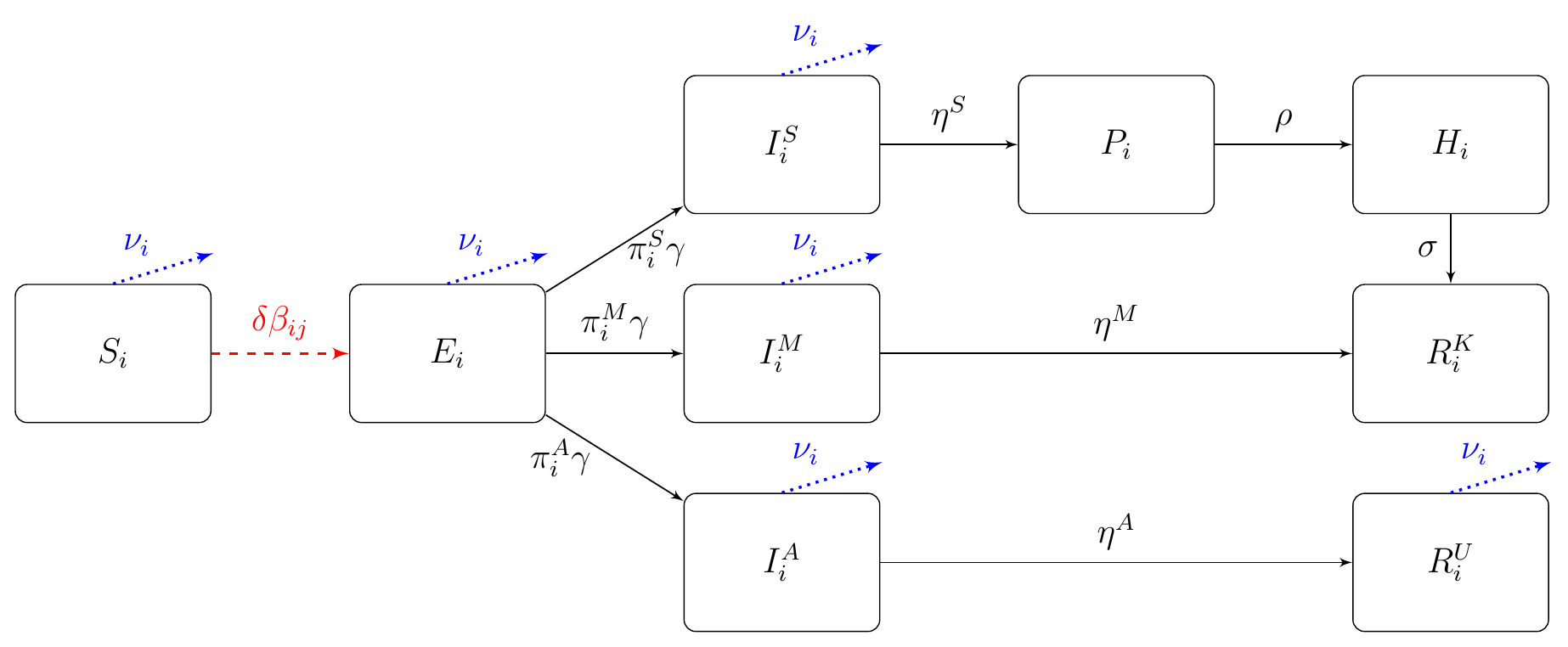}
	\caption{Non-vaccinated part.}
\end{subfigure}
\begin{subfigure}{0.9\textwidth}
	\centering
	\includegraphics[width=\textwidth]{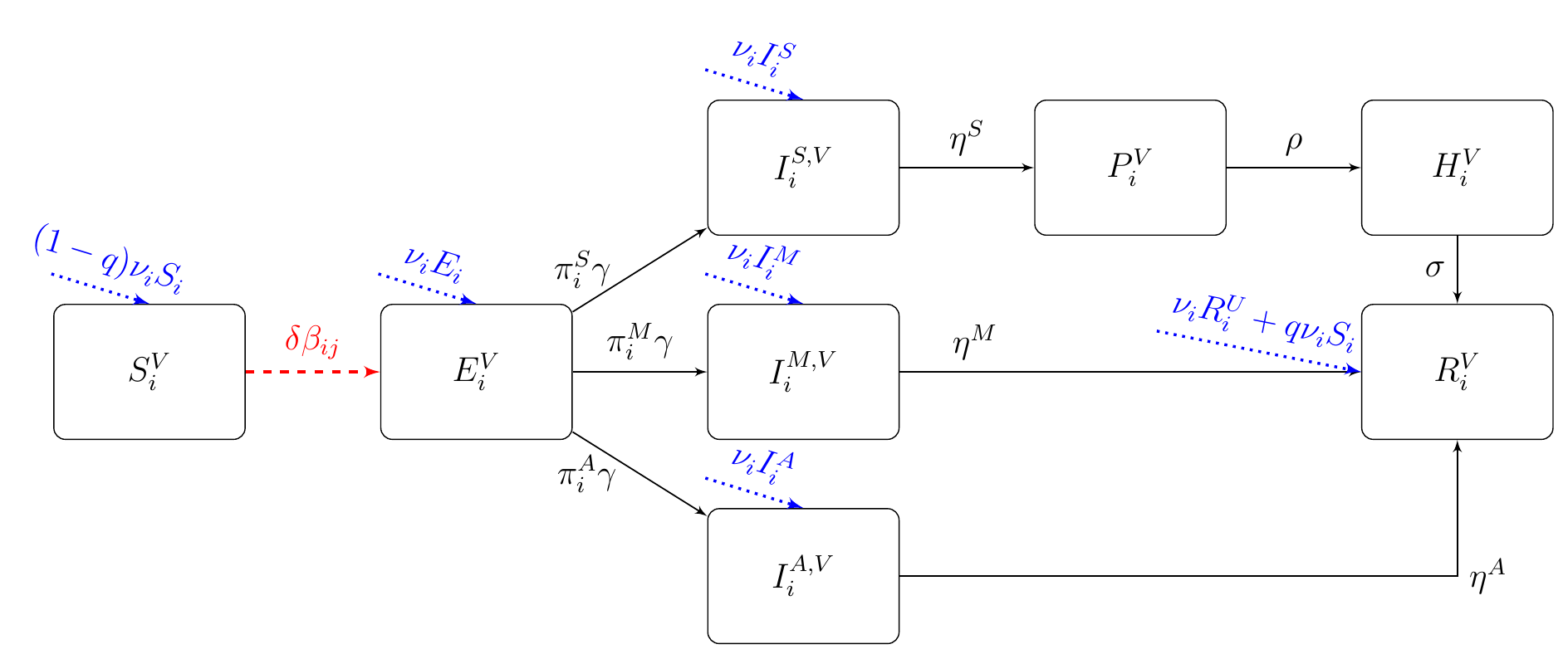}
	\caption{Vaccinated part.}
\end{subfigure}
	\caption{Flow of the SEIPHR model for the $i$-th age group. The controls associated with social distancing/vaccination are depicted with dashed red edges/dotted blue edges.}%
	\label{fig:flow_model}
\end{figure}

In the following, we describe the extended model, which is also shown in Figure~\ref{fig:flow_model}. %
In order to emphasize the effect of vaccination in combination with social distancing, we neglect mass testing in this paper and, thus, drop the $T$~compartments compared to~\cite{GrunHeyd20}. %
However, keeping Assumption~(A2) in mind, we split the SEIPHR model into two parts: The first one describes the compartments of people who have not been vaccinated while the second accounts for vaccinated people only.  For clarity the latter are marked by an additional superscript~$V$. Furthermore, we collect all infectious people in age group $i \in \{1,\ldots, n_\mathrm{g}\}$ at time $t \geq 0$ via
\begin{align}
	I_i(t) = I_i^S(t) + I_i^M(t) + I_i^A(t) + I_i^{S,V}(t) + I_i^{M,V}(t) + I_i^{A,V}(t). \notag
\end{align}

The non-vaccinated part of the dynamics reads as
\begin{subequations}\label{eq:model-1}
	\begin{align}
	\dot S_i(t) &  \; = \;  -\sum_{j=1}^{n_\mathrm{g}} \delta(t) \beta_{ij} S_i(t) I_j(t) - \nu_i(t) S_i(t) \label{eq:model-S} \\
	\dot E_i(t) &  \; = \;  \sum_{j=1}^{n_\mathrm{g}} \delta(t)\beta_{ij} S_i(t) I_j(t) - (\gamma + \nu_i(t)) E_i(t) \label{eq:model-E} \\
	\dot I_i^S(t) &  \; = \;  \pi_i^S \gamma E_i(t) - (\eta^S + \nu_i(t)) I_i^S(t) \\
	\dot I_i^M(t) &  \; = \;  \pi_i^M \gamma E_i(t) - (\eta^M + \nu_i(t)) I_i^M(t) \\
	\dot I_i^A(t) &  \; = \;  \pi_i^A \gamma E_i(t) - (\eta^A + \nu_i(t)) I_i^A(t) \\
	\dot R_i^U(t) &  \; = \;  \eta^A I_i^A(t) - \nu_i R_i^U(t) \\
	\dot P_i(t) &  \; = \; \eta^S I_i^S(t) - \rho P_i(t) \\
	\dot H_i(t) &  \; = \;  \rho P_i(t) - \sigma H_i(t) \\
	\dot R_i^K(t) &  \; = \;  \eta^M I_i^M(t) + \sigma H_i(t),
	\end{align}
\end{subequations}
where the control $\delta : [0,\infty[ \to [0,1]$ incorporates the average contact reduction as well as the transmission probability and $\nu_i : [0, \infty[ \to \mathbb{R}_{\geq 0}$ denotes the vaccination rate within age group~$i$. %
Following Assumption~(A1) we do not allow to vaccinate people who have been detected. %
Note that in our model exposed and infectious people are not yet detected. %
Those who get detected are immediately put into quarantine, i.e. they are either removed ($R^K$ and~$R^V$) or sent to the respective pre-ICU compartment ($P$ and~$P^V$); the others might be vaccinated. %
The parameter $1-q \in [0,1]$ describes the above mentioned probability of a vaccination failure, see Assumption~(A3).  %
From here on, we refer to $q$ as the \emph{success rate}. %
Furthermore,  compartments~$H$ and~$P$ collect all people in ICUs and those who have been detected, but do not yet require intensive care, respectively. %
    The parameters $ \beta = \left( \beta_{ij}\right)_{i,j = 1}^{n_\mathrm{g}} $ are the age-dependent transmission rates. These depend both on the amount of contacts between people of age group $ i $ and $ j $ as well as the probability that a contact of a susceptible and an infectious individual results in a transmission. %
    The inverse of~$\gamma$ is the mean incubation time, i.e. the time from infection to becoming infectious. %
    Furthermore, $\pi_i^S$, $\pi_i^M$, and~$\pi_i^A$ denote the age-dependent probabilities of having a severe, mild, or asymptomatic course of infection and
    $ \eta = (\eta^S, \eta^M, \eta^A) $ consists of the rates with which infectious individuals stop infecting susceptibles due to quarantine or recovery (including death). %
    Severe cases move from the pre-ICU compartments~$P_i$ to an ICU, i.e. compartment~$H_i$,  at rate~$\rho$ and are removed from this compartment at rate~$\sigma$. %
    For a more detailed description and interpretation of the parameters, we refer to~\cite{GrunHeyd20}; for the actual parameter values used see Table~\ref{tab:parameters} in Section~\ref{sec:results}.

The vaccinated part is then given by
\begin{subequations}\label{eq:model-2}
\begin{align}
	\dot S_i^V(t) & \; = \; (1-q) \nu_i(t) S_i(t) - \sum_{j=1}^{n_\mathrm{g}} \delta(t) \beta_{ij} S_i^V(t) I_j(t) \\
	\dot E_i^V(t) & \; = \; \nu_i(t) E_i(t) + \sum_{j=1}^{n_\mathrm{g}} \delta(t) \beta_{ij} S_i^V(t) I_j(t) - \gamma E_i^V(t) \\
	\dot I_i^{S,V}(t) & \; = \; \nu_i(t) I_i^S(t) + \pi_i^S \gamma E_i^V(t) - \eta^S I_i^{S,V}(t) \\
	\dot I_i^{M,V}(t) & \; = \; \nu_i(t) I_i^M(t) + \pi_i^M \gamma E_i^V(t) - \eta^M I_i^{M,V}(t) \\
	\dot I_i^{A,V}(t) & \; = \; \nu_i(t) I_i^A(t) + \pi_i^A \gamma E_i^V(t) - \eta^A I_i^{A,V}(t) \\
	\dot P_i^V(t) & \; = \;  \eta^S I_i^{S,V}(t) - \rho P_i^V(t) \\
	\dot H_i^V(t) & \; = \;  \rho P_i^V(t) - \sigma H_i^V(t) \\
	\dot R_i^V(t) & \; = \; \nu_i(t) R_i^U(t) + q \nu_i(t) S_i(t) + \eta^A I_i^{A,V}(t) + \eta^M I_i^{M,V}(t) + \sigma H_i^V(t),
\end{align}
\end{subequations}
where the transmission is subject to Assumptions~(A1) and~(A4). %
Note that we combine $R_i^V = R_i^{K,V} + R_i^{U,V}$ since we do not need to distinguish between known and unknown removed cases once they are vaccinated. %

We emphasize that~$q$ does not represent the efficacy often mentioned in media, as e.g.~\cite{Par20}. %
The latter considers two test groups -- one which gets the vaccine and another which gets a placebo. Then, only the patients who show symptoms are tested and from these, one infers the efficacy~\cite{Pfizer_on_efficacy}. %
However, in our model, the parameter~$q$ describes the probability of the patient being immune after the vaccination. %

In this paper, we focus on age-differentiated vaccination in combination with homogeneous contact restrictions among the age groups. %
The extension to heterogeneous social distancing is straightforward as elaborated in~\cite{GrunHeyd20}. %

For a concise notation, we collect all states in $x(t) \in \mathbb{R}^n$, controls in~$u(t) \in \mathbb{R}^m$, and parameters in $p \in \mathbb{R}^\ell$ and write
\begin{align}
	\dot x(t) \; = \; f(x(t),u(t),p), \quad x(0) = x^0, \label{eq:model-abstract}
\end{align}
with initial value~$x^0$.  Furthermore, all compartments describe fractions of the total population, i.e. $\sum_{i=1}^n x_i(t) = 1$ for all $t \geq 0$, and the proportion of age group~$i$ is denoted by~$N_i$, $i \in \{1,\ldots, n_\mathrm{g}\}$.  %
We assume that the countermeasures are kept constant over one week. %
Thus, the control~$u$ is piecewise constant which ensures existence and uniqueness of the solution of~\eqref{eq:model-abstract}. %
Moreover, a brief discussion on equilibria of system~\eqref{eq:model-1}--\eqref{eq:model-2} is given in Appendix~\ref{app:equilibria}. %

\section{Optimal Vaccination Strategy}\label{sec:optimal_control}
\begin{figure}[htbp!]
\centering
\includegraphics[scale=0.4]{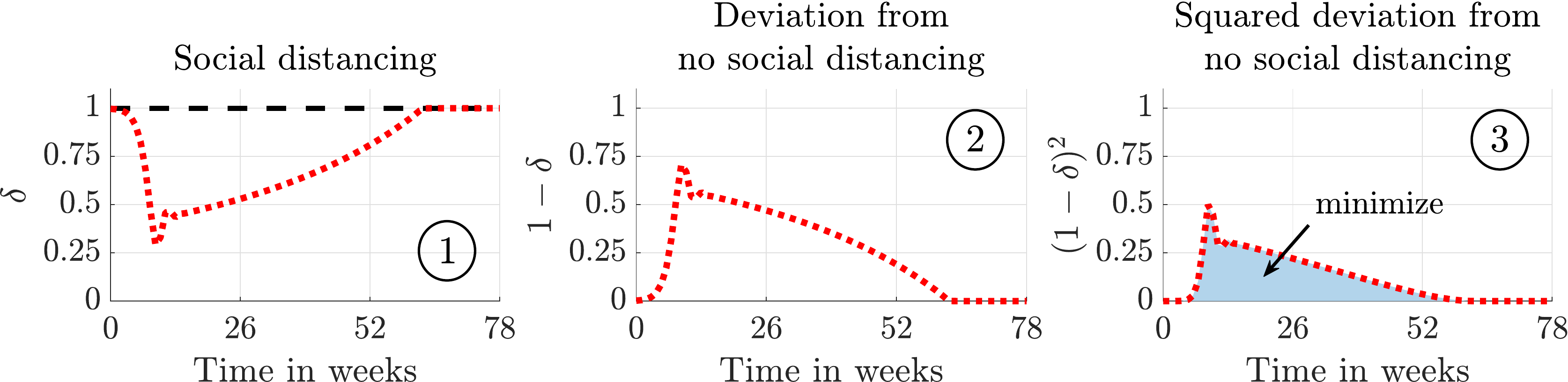} %
\caption{The three steps of evaluating the objective function. The dotted red line denotes $\delta$ (left), $1 - \delta$ (middle), and $(1 - \delta)^2$ (right) for a given social distancing profile, $\delta$. The dashed black line corresponds to no social distancing and the objective is to minimize the blue shaded area. We minimize the area under the squared deviation in order to discourage very strict contact restrictions.}
\label{fig:integral}
\end{figure}

In this section, we formulate an OCP to determine a coordinated social distancing and vaccination strategy that reduces the required social distancing while maintaining an ICU cap. %
To this end, we assume an amount of~$V^\mathrm{max}$, $V^\mathrm{max} \in \mathbb{N}$,  units of the vaccine to become available each day,  i.e. at time $t \geq 0$, the vaccine distribution is subject to
\begin{align*}
	n_\mathrm{pop} \cdot \int_0^t{ \sum_{i=1}^{n_\mathrm{g}} \nu_i(s) V_i(s) \, \mathrm{d} s} \leq V^\mathrm{max} t,
\end{align*}
where
\begin{align}
	V_i(s) = S_i(s) + E_i(s) + I_i^S(s) + I_i^M(s) + I_i^A(s) + R_i^U(s) \notag
\end{align}
collects all people in age group $i$ available for vaccination at time instant~$s$, and $n_\mathrm{pop} \in \mathbb{N}$ denotes the total population. %
Furthermore, we penalize social distancing by minimizing the objective function
\begin{align}
	J(\delta) = \int_{0}^{t_f}{ (1 - \delta(t))^2 \, \mathrm{d} t}. \notag
\end{align}
Figure~\ref{fig:integral} provides some intuition for~$J$. %
The OCP is then given by
\begin{subequations}\label{delta_problem}
\begin{align}
	\min_{(\delta, \nu)} \quad & J(\delta) + \kappa \left\| \nu \right\|_2^2 \label{eq:delta_obj_fcn} \\
	\text{subject to} \quad & n_\mathrm{pop} \cdot \sum_{i=1}^{n_\mathrm{g}} H_i(t) + H_i^V(t) \leq H^\mathrm{max} \label{eq:delta_constraint_Hmax} \\
	& \dot{x}(t) = f(x(t), u(t), p), \quad x(0) = x^0 \\
	& \delta(t) \in [0,1] \\
	& n_\mathrm{pop} \cdot \int_0^t{ \sum_{i=1}^{n_\mathrm{g}} \nu_i(s) V_i(s) \, \mathrm{d} s} \leq V^\mathrm{max} \cdot t, \quad \forall \, t \geq 0 \label{eq:constraint_Vmax} \\
	&\delta(t) = \delta(k\Delta t), \quad t\in[k\Delta t, (k+1)\Delta t[, \quad k = 0, \ldots, K-1 \label{eq:zoh:delta} \\
	&\nu(t)    = \nu(k\Delta t),    \quad t\in[k\Delta t, (k+1)\Delta t[, \quad k = 0, \ldots, K-1 \label{eq:zoh:nu}
\end{align}
\end{subequations}
where constraint~\eqref{eq:delta_constraint_Hmax} caps the total number of required ICU beds.  %
The positive parameter~$\kappa \ll 1$ weights the regularization term, which ensures that people are only vaccinated at points in time where it affects the objective function or inequality constraints. For instance, towards the end of the horizon, the number of susceptible people may become so low that the epidemic can no longer sustain itself even in the absence of social distancing measures. In this case, vaccination is not necessary. %
In~\eqref{eq:zoh:delta}-\eqref{eq:zoh:nu}, we assume the controls to be constant over one week, i.e., $\Delta t$ is one week, and $K$ is the total number of weeks considered. %

\begin{remark}
Note that the optimal solution of~\eqref{delta_problem} strongly depends on the objective function~\eqref{eq:delta_obj_fcn}. %
Different objectives yield different strategies. %
Here, we focus on reducing contact restrictions and, thus, economic as well as psychological damage. %
Implicitly, we bound the number of fatalities by enforcing the hard cap~\eqref{eq:delta_constraint_Hmax}  on the number of required ICUs. %
In Appendix~\ref{app:hosp_min} we additionally investigate how to minimize the total number of fatalities while keeping a constant level of social distancing. %
\end{remark}

\section{Case Study}\label{sec:results}
In this case study, we investigate
\begin{enumerate}
	\item the importance of vaccine \emph{availability} and \emph{efficacy},
	\item \emph{whom} to vaccinate first,
	\item how much we can \emph{relax} social distancing while distributing the vaccines,
	\item the difference between short- and long-term planning
\end{enumerate}
by analyzing numerical solutions to the OCP~\eqref{delta_problem}. %
We solve the OCPs using a direct single-shooting approach~\cite{binder2001introduction} combined with the standard sequential quadratic programming~\cite{Nocedal2006} algorithm implemented in \texttt{fmincon} in \texttt{Matlab}. Furthermore, we compute the gradients of the left-hand sides of the nonlinear inequality constraints~\eqref{eq:delta_constraint_Hmax} and~\eqref{eq:constraint_Vmax} using a continuous forward method (i.e. we numerically integrate the sensitivity equations forward in time). %
For convenience, we refer to contact reductions by up to \mbox{20\%} as \emph{light}, between \mbox{20\%} and \mbox{60\%} as \emph{strict}, and we consider reductions by over \mbox{60\%} a \emph{lockdown}. %

First, we consider long-term open-loop solutions. %
To this end, we develop time-varying strategies over the entire time span of the pandemic (approximately 2~years) at once. %
Specifically, we compare solutions for different values of the vaccination success rate, $q$, the number of vaccines supplied each day, $V^{\max}$, and the ICU capacity, $H^{\max}$.

Note that all simulations come along with uncertainties, e.g. resulting from not modelled effects, inaccurate parameters, new developments. %
This model-plant mismatch particularly causes problems for long-term simulations. %
In the context of mitigation of COVID-19 it is essential to update model parameters based on newly acquired data in order to develop adequate interventions. %
To this end, we also analyse short- to medium-term closed-loop solutions (where the strategies are updated repeatedly on newly available measurements). This emulates the real-life decision process where mitigation strategies are continuously updated when new data becomes available.

Throughout the simulations, we use the (fixed) parameters shown in Table~\ref{tab:parameters}. %
\begin{table}[htbp!]
	\centering
	\caption{Parameter values and references. For more details on the parameters see also \cite[Section 3]{GrunHeyd20}.}
	\renewcommand{\arraystretch}{1.05}
	\begin{tabular}{l|c|ccc|r}
		\hline
	\multicolumn{1}{l}{Description} & \multicolumn{1}{c}{Symbol} && Value & \multicolumn{1}{r}{} & Reference \\
		\hline
		Total population 	 & $n_\mathrm{pop}$ && $8.3 \cdot 10^7$ & &\cite{destatis20b}\\
		Number of age groups & $n_\mathrm{g}$   && $3$  && --\\
		Regularization parameter & $\kappa$ && $10^{-3}$ && --\\
		Removal rate (severe) & $\eta^S$ && 0.2500 && \cite{GrunHeyd20,he_temporal_2020} \\
		Removal rate (mild) & $\eta^M$ && 0.2500 && \cite{GrunHeyd20,he_temporal_2020}\\
		Removal rate (asymptomatic) & $\eta^A$ && 0.1667 && \cite{woelfel_clinical_2020}\\
		Rate of becoming infectious & $\gamma$ && 0.1923 && \cite{lauer_incubation_2020}\\
		ICU admittance rate & $\rho$   && 0.0910 && \cite{dreher_charakteristik_2020}\\
		ICU discharge rate  & $\sigma$ && 0.0952 && \cite{dreher_charakteristik_2020}\\
		 \hline
		 \multicolumn{6}{c}{Age-differentiated parameters} \\
		 \hline
		\multicolumn{1}{l}{Age group} & \multicolumn{1}{c}{$i$} & 1 & 2 & \multicolumn{1}{r}{3} & Reference\\
		 \hline
		 Age range (in years)  			& 		--		& $<$ 15 & 15 -- 59 & $\geq$ 60 & --\\
		 Relative age group size         & $N_i$        & 0.1370 & 0.5776 & 0.2854 & \cite{destatis20b} \\
		 Probability of severe symptoms  & $\pi_i^S$    & 0.0053 & 0.0031 & 0.0302 & \cite{schilling_vorlaufige_2020}\\
		 Probability of mild symptoms    & $\pi_i^M$    & 0.1211 & 0.2201 & 0.2512 & \cite{schilling_vorlaufige_2020}\\
		 Probability of no symptoms      & $\pi_i^A$    & 0.8737 & 0.7768 & 0.7186 & \cite{schilling_vorlaufige_2020}\\
		 Transmission rate (age group 1) & $\beta_{1i}$ & 0.4612 &        &        & \cite{MossHens08,park_systematic_2020}\\
		 Transmission rate (age group 2) & $\beta_{2i}$ & 0.4819 & 0.6304 &        & \cite{MossHens08,park_systematic_2020}\\
		 Transmission rate (age group 3) & $\beta_{3i}$ & 0.1243 & 0.2944 & 0.1802 & \cite{MossHens08,park_systematic_2020}\\
		 \hline
	\end{tabular}
	\label{tab:parameters}
\end{table}

\subsection{Long-Term Simulations}\label{sec:open:loop}
\begin{figure}[htbp!]
	\centering
	\includegraphics[scale=0.4]{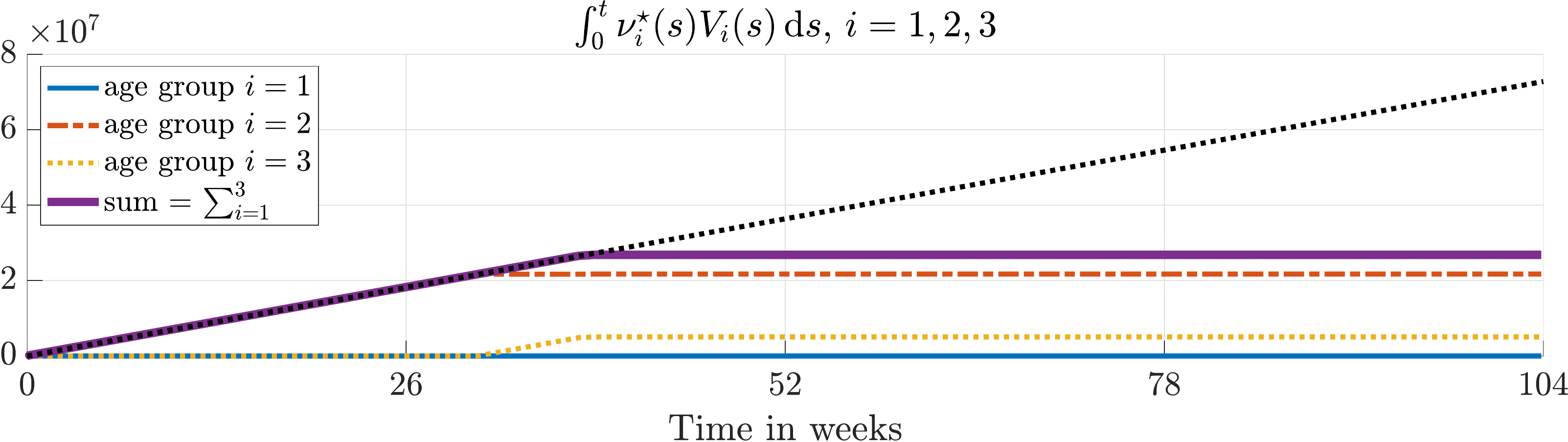}
\caption{Optimal vaccination strategy $\nu^\star = (\nu_1^\star, \nu_2^\star, \nu_3^\star)$ with vaccination success rate $q = 0.9$ and daily available units $V^\mathrm{max} = 10^5$; the dotted black line depicts~$V^\mathrm{max} t$. The vaccination process is stopped once all contact restrictions are lifted (Figure~\ref{fig:res_delta_H}) and the pandemic is contained without further interventions, i.e. after approximately 40~weeks.}
\label{fig:res_vaccination}
\end{figure}

\begin{figure}
	\centering
	\includegraphics[scale=0.4]{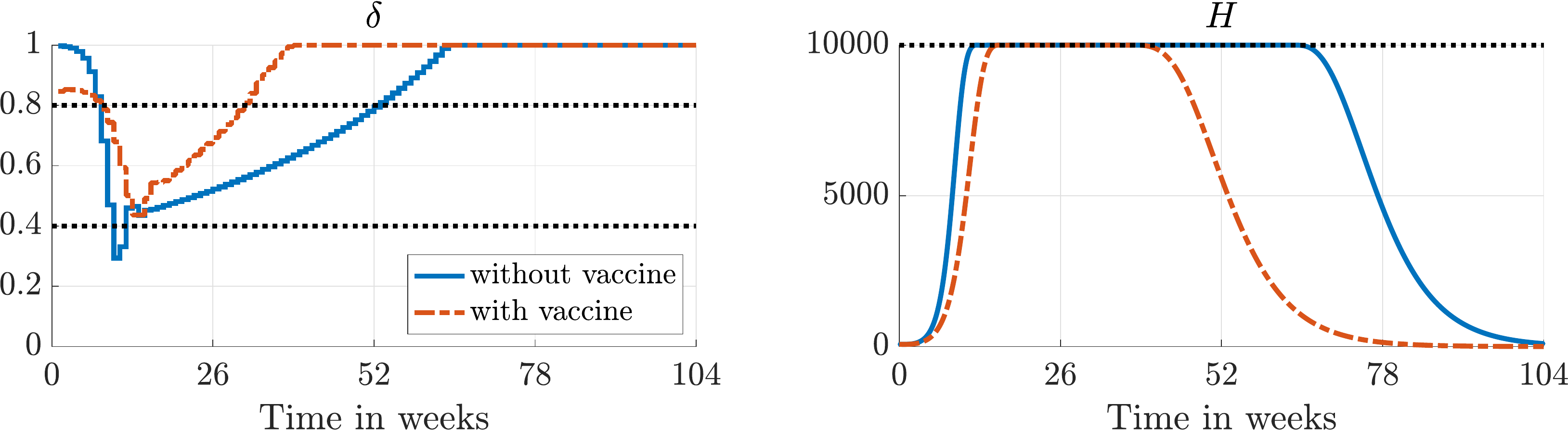}
	\caption{Optimal social distancing profile~$\delta$ (left) and ICU occupancy $H = \sum_{i=1}^{n_\mathrm{g}} H_i + H_i^V$ (right) with and without vaccination. The dotted black lines mark strict social distancing, i.e. $\delta \in [0.4, 0.8]$ (left) and~$H^\mathrm{max}$ (right). }
	\label{fig:res_delta_H}
\end{figure}

Results for the OCP~\eqref{delta_problem} can be found in Figures~\ref{fig:res_vaccination} and~\ref{fig:res_delta_H}. %
Here, we chose $q = 0.9$ since the general expectation is that the success rate is quite high. Furthermore, $V^\mathrm{max} = 100,000$ and $H^\mathrm{max} = 10,000$ are based on~\cite{Spiegel20, divi_october_2020}. %
The vaccination constraint~\eqref{eq:constraint_Vmax} is active until the contact restrictions are lifted. %
At that point, the pandemic is contained without further interventions. %
Our solution suggests to not vaccinate the high-risk group first, but rather the middle-aged group. %
The objective of this is to minimize social distancing. Specifically, it allows us to relax the social distancing measures earlier while still maintaining the hard infection cap modelled by the upper bound on the ICU capacity. Hence, the elderly (high-risk group) get vaccinated once the social distancing is significantly reduced, i.e. around week~36. %
In contrast, we demonstrate in Appendix~\ref{app:hosp_min} that if the objective is to minimize fatalities, it is optimal to vaccinate the elderly first provided that the social distancing measures are not too strict. However, in this case, the ICU occupancy exceeds the capacity. %

Furthermore, when no vaccine is available, an actual lockdown is optimal. %
Without vaccination, strict social distancing is required for 43~weeks. With vaccination, the strict social distancing can be lifted approximately 20~weeks earlier, i.e. a reduction by almost \mbox{50\%}. The same conclusion applies to the ICU occupancies. %

\begin{figure}[htbp!]
\centering
\includegraphics[scale=0.4]{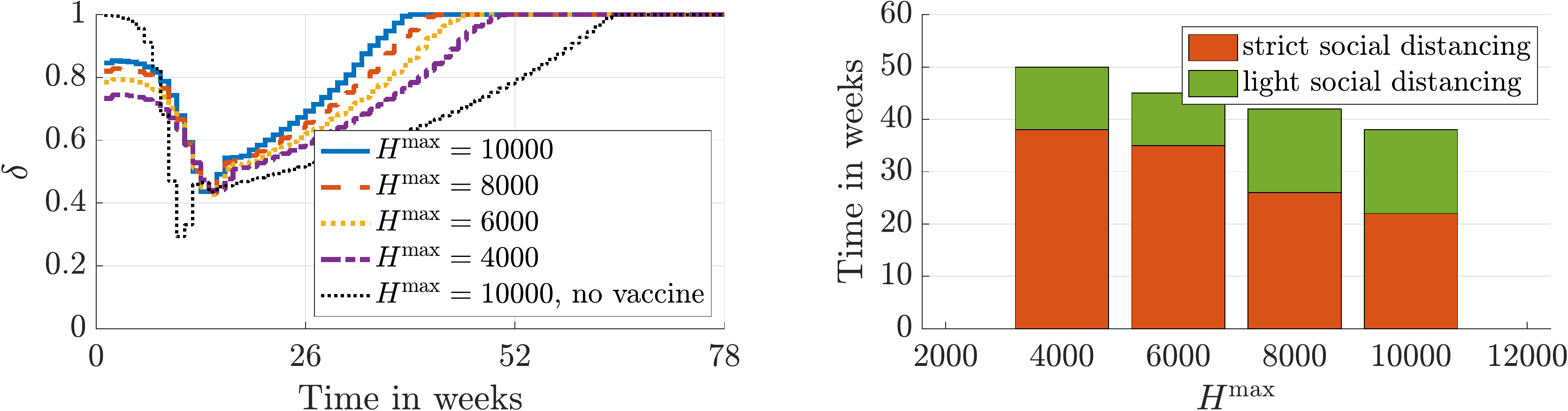}
\caption{Impact of~$H^\mathrm{max}$, i.e. the number of available ICUs, on the required social distancing with vaccination success rate $q = 0.9$ and daily available units of vaccines $V^\mathrm{max} = 10^5$.}
\label{fig:impact_H}
\end{figure}

Figure~\ref{fig:impact_H} (left) shows the optimal social distancing strategies for different ICU capacities, $H^{\max}$.
Interestingly, the maximum amount of contact restrictions, $\delta = 0.435$, is (almost) independent of~$H^\mathrm{max}$. %
Consequently, it is possible to reduce the total number of people who become admitted to ICUs at the cost of longer but not stricter social distancing measures. %
This is particularly important as the number of people in ICUs is closely related to the number of fatalities. %
Figure~\ref{fig:impact_H} (right) visualizes the impact of the number of available ICUs on the number of weeks where contact restrictions have to be enforced.

\begin{figure}
\centering
\begin{minipage}[t]{0.55\textwidth}
\vspace{0pt}
\includegraphics[scale=0.4]{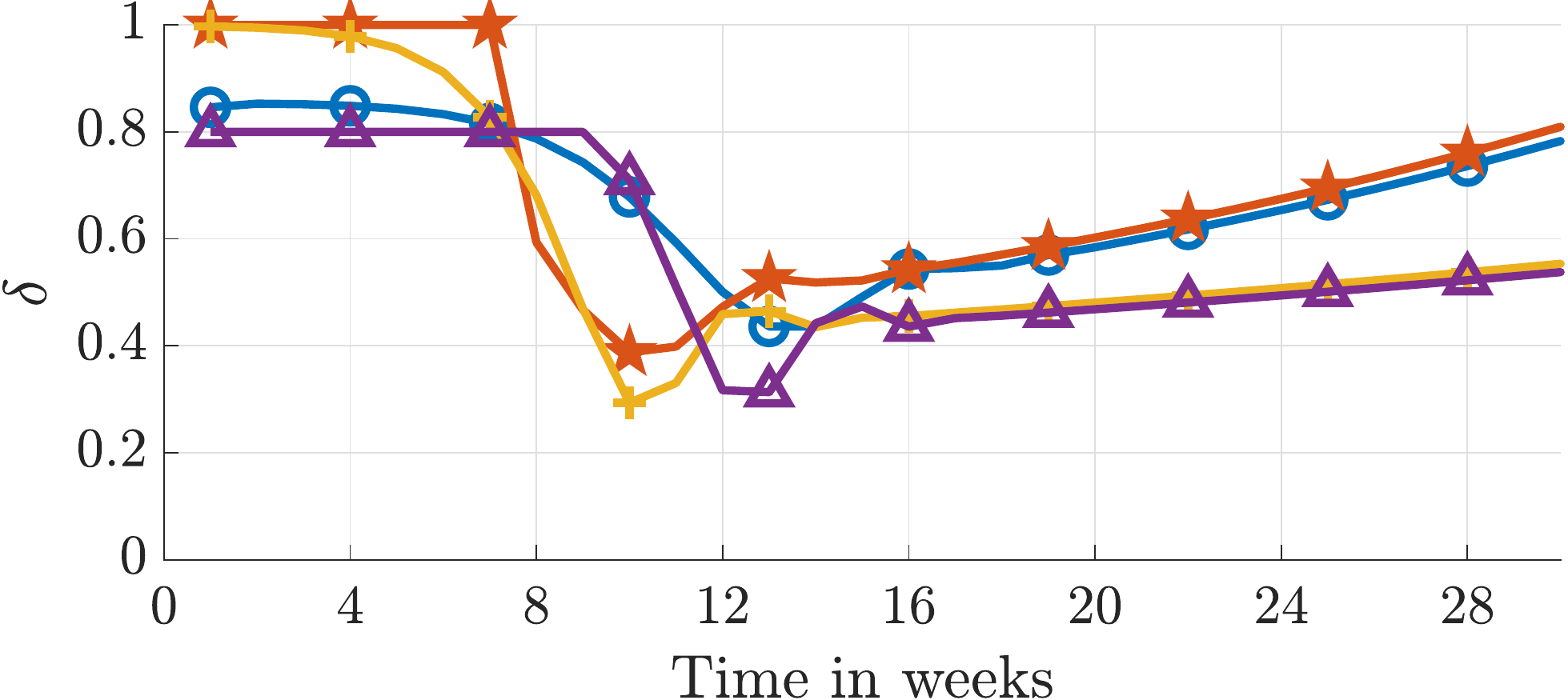}
\end{minipage}
\begin{minipage}[t]{0.4\textwidth}
\begin{scriptsize}
\vspace{0pt}
\begin{tabular}{c|lc}
  & explanation & $J(\delta)$ \\
\hline
${\color{matlabblue}{\circ}}$ & vaccine & 13.25 \\
${\color{matlabred}\star}$ & vaccine, prohibiting & 14.03 \\
${\color{matlabyellow}+}$ & no vaccine & 31.47 \\
${\color{matlabpurple}\triangle}$ & no vaccine, enforcing & 32.34
\end{tabular}
\end{scriptsize}
\end{minipage}
\caption{Impact of enforcing/prohibiting social distancing during the initial phase with $H^\mathrm{max} = 10^4$ on the optimal social distancing strategy~$\delta$ (left) and total amount, $J(\delta)$, of contact restrictions (right).}
\label{fig:initial_phase}
\end{figure}

Moreover, the solutions with vaccination all enforce contact restrictions from the beginning whereas the solution without vaccination lets the pandemic evolve for some weeks before implementing a strict lockdown. %
We explain this using Figure~\ref{fig:initial_phase} which shows the optimal social distancing strategies with~(${\color{matlabblue}\circ}$) and without~(${\color{matlabyellow}+}$) vaccination (also shown in Figure~\ref{fig:res_delta_H}). We compare these two strategies to 1) enforcing contact restrictions in the beginning ($\delta=0.8$) without any vaccination~(${\color{matlabpurple}\triangle}$) and 2) prohibiting them in the beginning ($\delta=1$) and allowing vaccination~(${\color{matlabred}\star}$).
The key observation is that without social distancing in the beginning, a hard lockdown is necessary regardless of whether a vaccine is available or not. However, when a vaccine is available, some social distancing in the beginning avoids a harder lockdown later because a significant amount of people are already vaccinated at this point. Furthermore, if no vaccine is available, it is not beneficial to enforce social distancing early on because it only slows down the \emph{natural} vaccination (i.e. the immunity following an infection).

\begin{figure}[htbp!]
\centering
\includegraphics[scale=0.4]{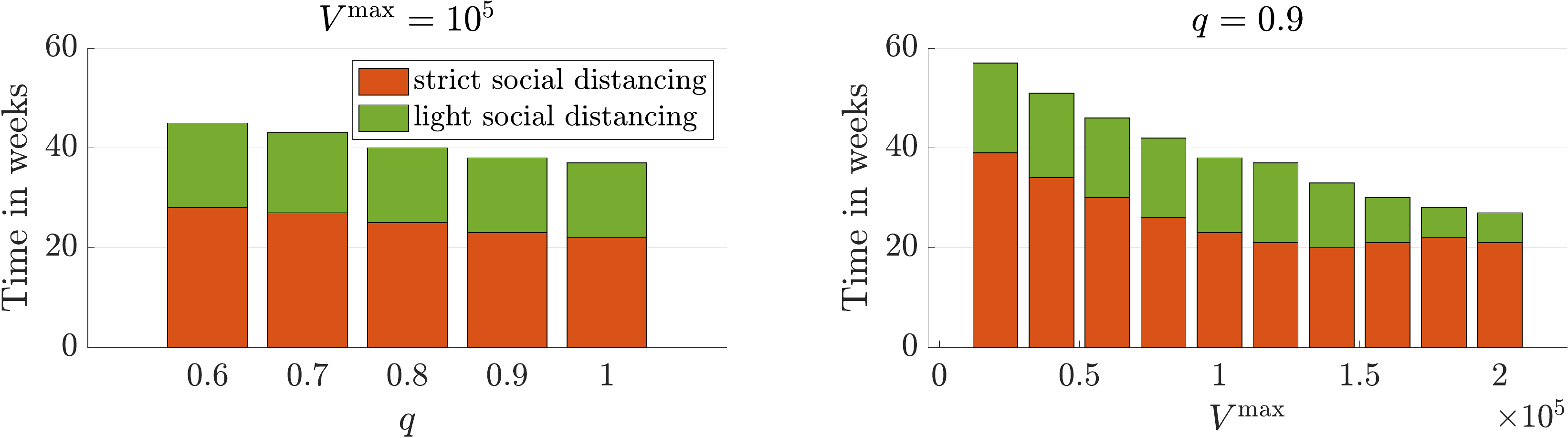}
\caption{Impact of vaccination success rate~$q$ (left) and daily production rate~$V^\mathrm{max}$ (right) on the total time contact restrictions have to be implemented with ICU cap $H^\mathrm{max} = 10^4$.}
\label{fig:impact_q_Vmax_bar}
\end{figure}
\begin{figure}
\centering
\includegraphics[scale=0.4]{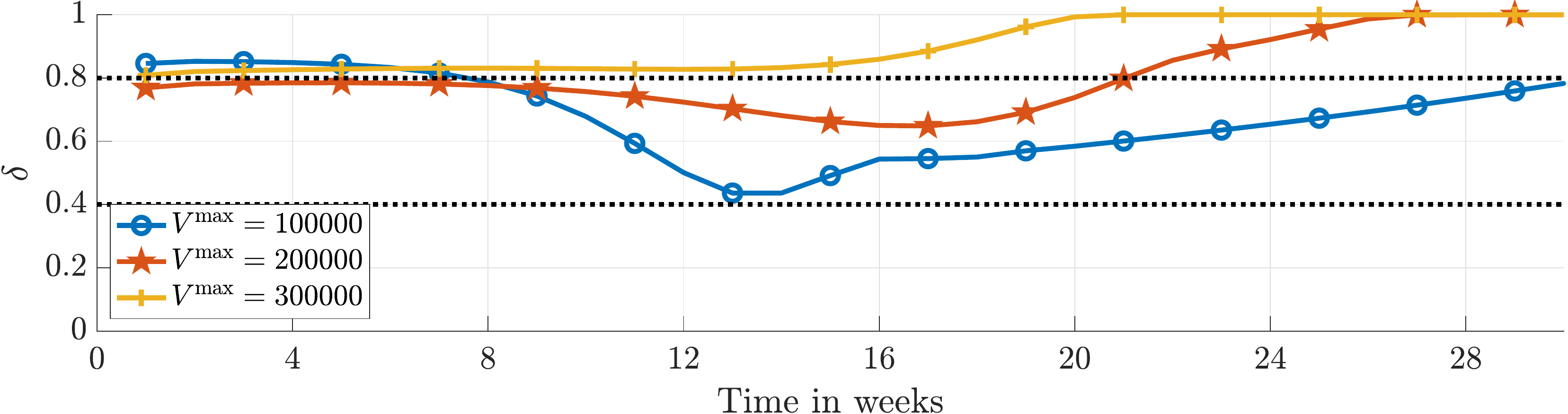}
\caption{Impact of the amount of daily available units of vaccine~$V^\mathrm{max}$ on the optimal social distancing strategy~$\delta$ with vaccination success rate $q = 0.9$ and ICU cap $H^\mathrm{max} = 10^4$.}
\label{fig:impact_largeV}
\end{figure}
According to~\cite{Spiegel20}, a realistic number of daily vaccines is $V^\mathrm{max} = 100,000$. %
However, at the time of submitting this manuscript, no actual numbers are available. %
For this reason, we investigate the impact of varying both the available number of vaccines as well as the success rate on the social distancing and on the end time of the contact restrictions.  %
The results depicted in Figure~\ref{fig:impact_q_Vmax_bar} show that it is more important to increase the available amount of the vaccines than the success rate. %
Note that when increasing~$V^\mathrm{max}$ (right), there is a threshold where the required amount of strict social distancing increases before it decreases again. %
The reason can be seen in Figure~\ref{fig:impact_largeV}. For $V^{\max} = 200,000$, the contact reductions become classified as strict in the beginning whereas for lower and higher values, the contact reductions are still classified as light. %

\begin{figure}
\centering
\includegraphics[scale=0.4]{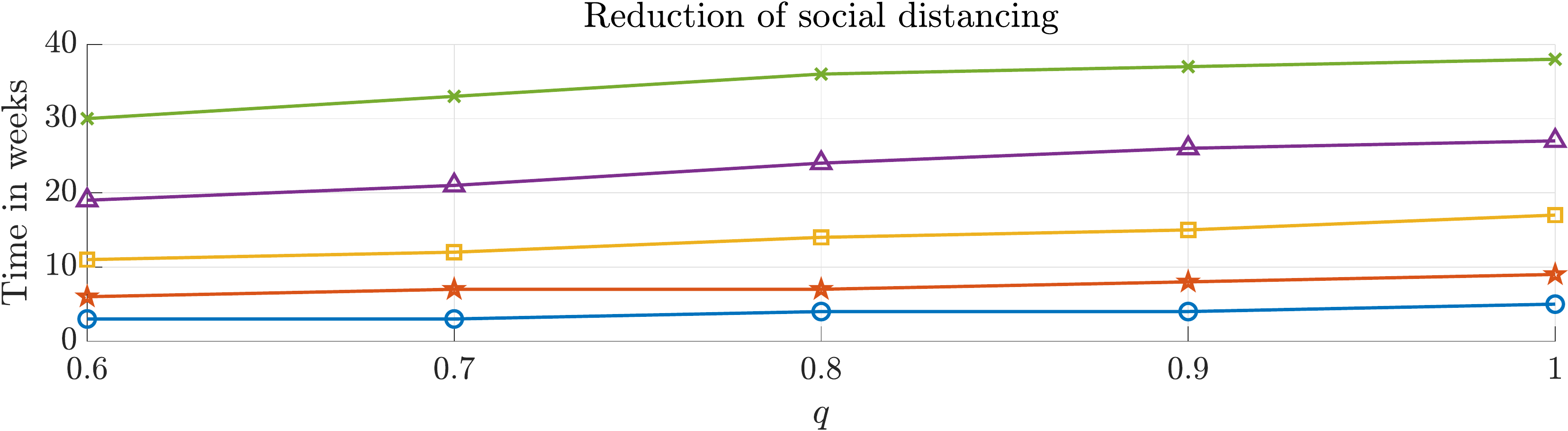}
\caption{Impact of the amount of vaccination success rate~$q$ and daily available units of vaccine~$V^\mathrm{max}$ on the time until all restrictions are relaxed compared to the case without vaccine. Different markers denote different amounts of vaccine~$V^\mathrm{max}$: ${\color{matlabblue}\circ} = 1.25 \cdot 10^4$, ${\color{matlabred}\star} = 2.5 \cdot 10^4$, ${\color{matlabyellow}\Box} = 5 \cdot 10^4$, ${\color{matlabpurple}\triangle} = 10^5$, ${\color{matlabgreen}\times} = 2 \cdot 10^5$. The ICU cap is set to $H^\mathrm{max} = 10^4$.}
\label{fig:impact_q_Vmax}
\end{figure}

Furthermore, Figure~\ref{fig:impact_q_Vmax} provides some insight on how to improve the impact of a vaccine. For instance, if the supply rate is small, e.g. $V^\mathrm{max} = 12,500$, improving the success rate~$q$ does not help reducing the contact restrictions. %
Once sufficiently many units of vaccine can be produced, it is possible to reduce the social distancing by increasing the success rate.

Based on the results shown in Figures~\ref{fig:impact_q_Vmax_bar} and \ref{fig:impact_q_Vmax}, we conclude that increasing the number of \emph{successful} vaccinations has the biggest impact on the social distancing profile. For instance, if $q = 0.5$ and $V^\mathrm{max} = 100,000$, doubling~$V^\mathrm{max}$ has a bigger impact than increasing $q$ to $0.9$. Doubling $V^\mathrm{max}$ adds another $50,000$ successful vaccinations whereas increasing~$q$ only adds another $40,000$ successful vaccinations.

\subsection{Consecutive Short-Term Simulations}\label{sec:mpc}
In order to simulate real-life decision-making processes and to account for uncertainties as mentioned in the beginning of Section~\ref{sec:results}, we use MPC~\cite{CoroGrue20}. %
The main idea of MPC is to solve a sequence of OCPs of the form 
\begin{subequations}\label{mpc_problem}
\begin{align}
	\min_{(\delta, \nu)} \quad & \int_{k \Delta t}^{(k+K) \Delta t}{(1 - \delta(t))^2 \, \mathrm{d} t} + \kappa \left\| \nu \right\|_2^2 \\
	\text{subject to} \quad & n_\mathrm{pop} \cdot \sum_{i=1}^{n_\mathrm{g}} H_i(t) + H_i^V(t) \leq H^\mathrm{max} \\
	& \dot{x}(t) = f(x(t), u(t), p), \quad x(0) = x^0 \\
	& \delta(t) \in [0,1] \\
	& n_\mathrm{pop} \cdot \int_{k \Delta t}^t{ \sum_{i=1}^{n_\mathrm{g}} \nu_i(s) V_i(s) \, \mathrm{d} s} \leq V^\mathrm{max} \cdot (t - k \Delta t) + V^k \\
	&  \forall \, t \in [k \Delta t, (k+K) \Delta t] \\
	& \delta(t) = \delta(k \Delta t), \quad t \in [k\Delta t, (k+1)\Delta t[, \quad k = 0, \ldots, K-1 \\
	& \nu(t)    = \nu(k \Delta t),    \quad t \in [k\Delta t, (k+1)\Delta t[, \quad k = 0, \ldots, K-1
\end{align}
\end{subequations}
over a moving time window of length $K \Delta t$, where $K \in \mathbb{N}_{\geq 2}$ denotes the number of time steps of length $\Delta t > 0$. %
Here, the parameter $V^k$ accounts for the units of vaccine that have been saved in previous MPC steps. %
This scheme can be summarized as follows. \\[-2mm]

\begin{center}
\begin{minipage}{0.95\textwidth}
{\bf for $k = 0$ to $\mathcal{K} - 1$}
\begin{enumerate}
	\item Measure/estimate current state $x(k \Delta t) = x^k$ at the $k$-th time instant.
	\item Solve the OCP~\eqref{mpc_problem} on the time window $[k \Delta t, (k+K) \Delta t]$ to get an optimal control $u^\star : [k \Delta t, (k+K)\Delta t] \to \mathbb{R}^m$.
	\item Implement the first portion of the solution $u^k|_{[k \Delta t, (k+1) \Delta t]} = u^\star|_{[k \Delta t, (k+1) \Delta t]}$.
	\item Increment $k \leftarrow k+1$.
\end{enumerate}
{\bf end}\\[-2mm] 
\end{minipage}
\end{center}
Here, $\mathcal{K} = t_f / \Delta t$ denotes the number of MPC steps, i.e. the number of OCPs of the form~\eqref{mpc_problem} that have to be solved to arrive at a solution of~\eqref{delta_problem}. In our simulations, we set $\Delta t$ to one week. %

\begin{figure}[htbp!]
\centering
\includegraphics[scale=0.4]{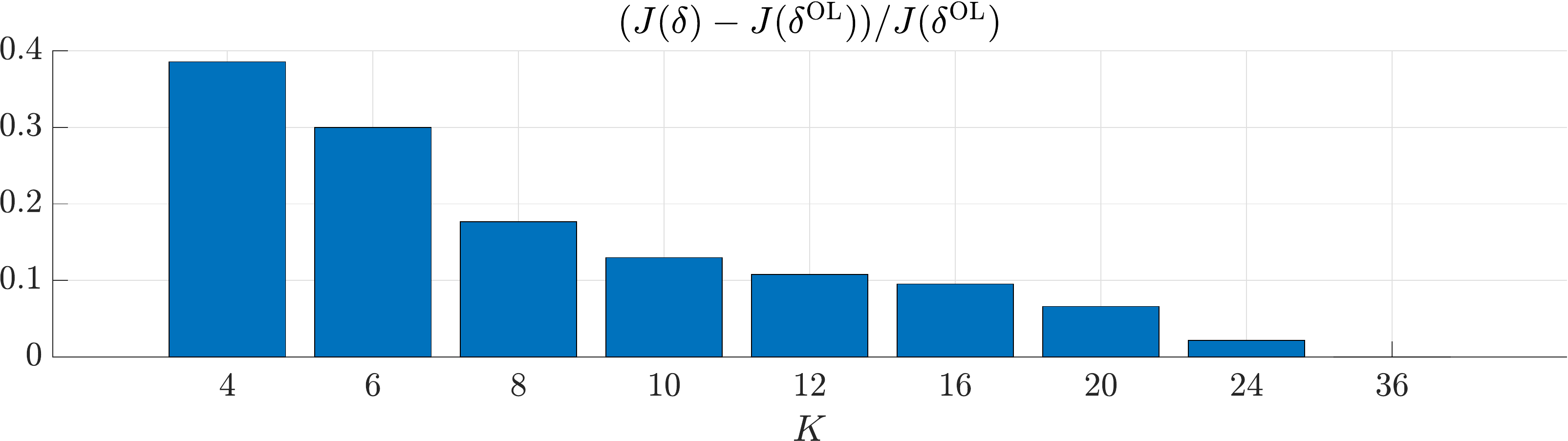}
\caption{Impact of prediction horizon length~$K$ in weeks on objective function value~$J(\delta)$, i.e. on the required amount of social distancing, compared to the open-loop (OL) solution. For $K < 3$ the ICU cap is violated (short-sightedness of MPC).}
\label{fig:cl_impact_N}
\end{figure}

We study the impact of different prediction horizon lengths~$K$ on the closed-loop solution. %
This corresponds to a simulation of the whole pandemic, while the decision for the next week is always made based on a forecast horizon of only $K$ weeks. %
It is essential to choose the prediction horizon sufficiently long; otherwise, the ICU caps might be violated due to time delays within the model, e.g. caused by the incubation time. %
In our simulations, the smallest possible integer value for the prediction horizon~$K$ that allows for maintaining the ICU cap is 3~weeks. %
The impact of the choice of $K \geq 4$ on the objective function value is visualized in Figure~\ref{fig:cl_impact_N}. %
The longer the prediction, the closer the objective function value gets to the one corresponding to the open-loop solution shown in the previous subsection. %
Moreover, the marginal gain of horizon lengths larger than eight weeks is negligible. %
Therefore, if a prediction horizon of eight weeks is used, the strategy does not suffer from the short-sightedness~\cite{DiBaDiPi18} of operating on a limited time window (but with significantly reduced uncertainty) while being able to adapt to newly acquired data. %

\begin{figure}[htbp!]
\centering
\includegraphics[scale=0.4]{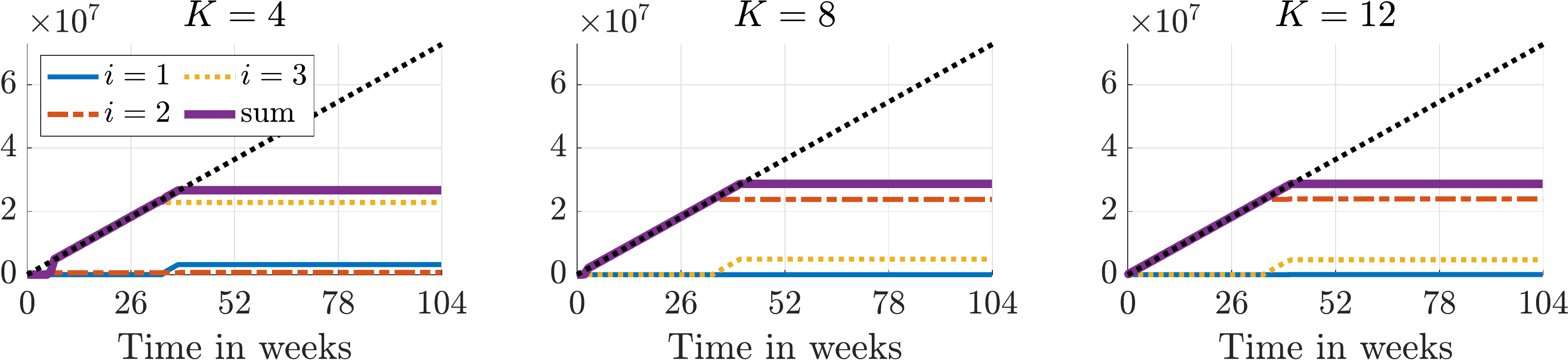}
\caption{Impact of prediction horizon length~$K$ on vaccination strategy. For short prediction horizons the high-risk groups are vaccinated first (immediate impact), for long prediction horizons the people with most contacts are vaccinated first in order to relax the social distancing in the long run.}
\label{fig:res_delta_cl_nuV}
\end{figure}

\begin{figure}[htbp!]
\centering
\includegraphics[scale=0.4]{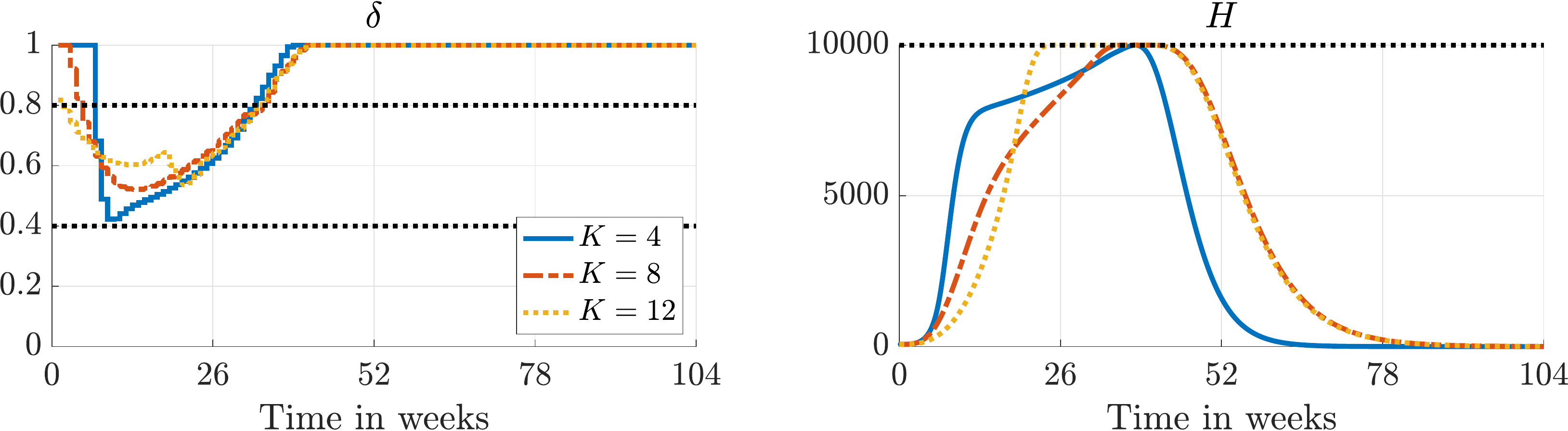}
\caption{Impact of prediction horizon length~$K$ on social distancing and ICU capacities. Short prediction horizons might yield necessity of a (short) lockdown.}
\label{fig:res_delta_cl_delta_H}
\end{figure}

The main difference compared to the MPC investigation in~\cite{GrunHeyd20} is that here the prediction horizon length crucially affects the optimal control. %
In particular, the choice of~$K$ is essential to answer the question \emph{whom to vaccinate first}. %
Optimal vaccination and social distancing strategies depending on the prediction horizon length are depicted in Figures~\ref{fig:res_delta_cl_nuV} and~\ref{fig:res_delta_cl_delta_H}, respectively. %
If the prediction horizon is small, it is optimal to vaccinate the high-risk group only (Figure~\ref{fig:res_delta_cl_nuV}), since they directly affect the number of required ICUs. %
Similar (open-loop) strategies are obtained if the number of fatalities is minimized (see Appendix~\ref{app:hosp_min}). %
Furthermore, the contact restrictions are severe, but are lifted comparatively early (Figure~\ref{fig:res_delta_cl_delta_H}). %
For longer prediction horizons, the people with the highest contact rates are vaccinated first, and the restrictions are not as severe. %
In particular, the vaccination strategies for $K = 8$ and $K = 12$ coincide. Furthermore, for these prediction horizons, the contact restrictions are not as strict as in the open-loop policy, they develop more smoothly, and they can be lifted monotonically. %
Note that with prediction horizon length $K = 12$ weeks, the social distancing around week~13 is quite relaxed ($\delta \geq 0.6$). As a result more restrictions have to be enforced around week~22. %

\section{Conclusions and outlook}\label{sec:conclusions}
In this paper, we address the four questions related to simultaneous vaccination and social distancing set out in the introduction. We address them by extending a previous compartmental model to include vaccination. Based on this model, we use optimal control and MPC to compute time-varying vaccination and social distancing	profiles which minimize the necessary amount of social distancing. %
Our simulations show that contact restrictions can be lifted almost half a year earlier as compared to a scenario without vaccination. This is achieved by first vaccinating the middle-aged group which is most responsible for spreading the virus. Thereafter, the elderly, which are most vulnerable to COVID-19, are vaccinated.
Furthermore, we find that the contact restrictions can be lifted even earlier by increasing the number of \emph{successful} vaccinations, which depends on both the number of available vaccines and the efficacy. We also observe that the maximal ICU occupancy can be reduced by extending the contact restrictions and that it is not necessary to make them more strict.
Additionally, we demonstrate that if the objective is to minimize the total number of fatalities and contacts are not sufficiently restricted, it is optimal to vaccinate the elderly first, see Appendix~\ref{app:hosp_min}. %
The above conclusions assume that long-term planning is possible, e.g. over a 2~year period. This is not possible in practice, but we demonstrate that optimal closed-loop profiles obtained with an 8~week prediction horizon are qualitatively similar to those obtained with a 2~year horizon. However, if too short horizons are used, only the elderly are vaccinated, and more social distancing is necessary.

In future work, we will account for the uncertainty of the model parameters and use \emph{uncertainty quantification} techniques to assess their impact on the policies presented in this paper.

\section*{Acknowledgments}
We thank Martin Wei{\ss}leder (BWK Berlin) for some insights on the medical process of vaccination. %

\appendix

\section{Equilibria}\label{app:equilibria}
In this appendix, we briefly discuss equilibria of the model presented in Section~\ref{sec:novel:seiphr:model}, i.e. we are interested in points $(\bar{x}, \bar{u}) \in \mathbb{R}^n \times \mathbb{R}^m$ with
\begin{align}
	f(\bar{x}, \bar{u}, p) \; = \; 0. \label{eq:app:steady_state}
\end{align} 
Furthermore, we address the relation between stability of an equilibrium and herd immunity, i.e., a point during the epidemic where sufficiently many people have become immune such that the spread dies out. %

\begin{figure}[htbp!]
\centering
\includegraphics[width=0.9\textwidth]{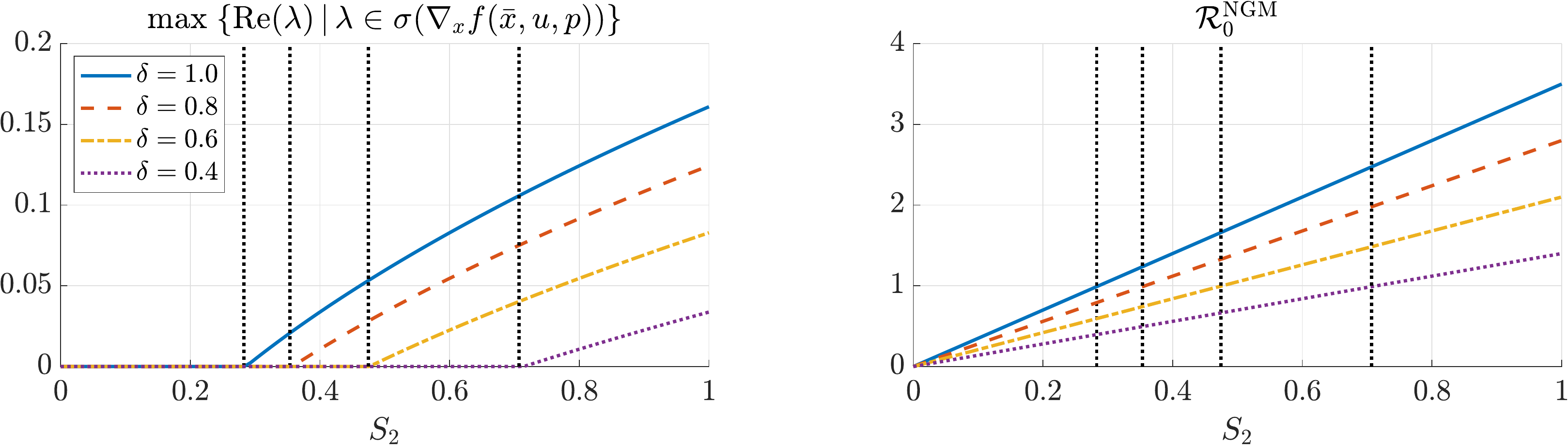}
\caption{Stability of equilibrium~$\bar{x}$ (left) and basic reproduction number~$\mathcal{R}_0^\mathrm{NGM}$ (right): If herd immunity has set in, i.e. $S = S_2$ is sufficiently small, the equilibrium point is stable; otherwise adding infected people might cause an outbreak. The vertical dotted black lines indicate herd immunity. By~$\sigma(A)$ we denote the spectrum of a matrix~$A$.}
\label{fig:stability_of_equilibrium}
\end{figure}

Herd immunity can be characterized in terms of the basic reproduction number~$\mathcal{R}_0^{NGM}$, i.e., the largest eigenvalue of the \emph{next generation matrix}, for details we refer~\cite{DiekHees90,DrieWatm02}. %
We say that \emph{herd immunity has set in if $\mathcal{R}_0^{NGM} < 1$}. %
The next generation matrix relates the inflow and the outflow of all infected compartments. %
Thus, it depends on both the state~$\bar{x}$ and the control~$\bar{u}$. %
In particular, for smaller~$\delta$, i.e., stricter social distancing and, hence, less inflow to the infected compartments, herd immunity sets in earlier, see also Figure~\ref{fig:stability_of_equilibrium} (right). %

To investigate equilibria, note that in our model people move unidirectionally from~$S_i$ to~$R_i$, i.e. either $R_i^U$, $R_i^K$, or~$R_i^V$, and stay there, see equations~\eqref{eq:model-1}--\eqref{eq:model-2} and Figure~\ref{fig:flow_model}. %
Moreover, since births and transitions among age groups are neglected and, hence, $S_i(t) \leq S_i(0)$ for all $t \geq 0$ and $i \in \{1, \ldots, n_\mathrm{g}\}$, condition~\eqref{eq:app:steady_state} can only be satisfied if there are no transitions among compartments. %
Consequently, there only exist disease-free equilibria, i.e., only $S_i$, $S_i^V$, $R_i^K$, $R_i^U$, and $R_i^V$ might be non-empty. %
Clearly, $(\bar{x}, \bar{u})$ with $\bar{x} = (S_1, \ldots, S_{n_\mathrm{g}}, 0_{n-n_\mathrm{g}})$ and $\sum_{i=1}^{n_\mathrm{g}} S_i = 1$ is a disease-free equilibrium. %
In the following we consider $\sum_{i=1}^{n_\mathrm{g}} S_i  < 1$ and distinguish three cases depending on the vaccination strategy. %
First, if people are vaccinated all the time, i.e., $\min_i \nu_i(t) > 0$ for all~$t$, then compartments~$S_i$ and~$R_i^U$ have to be zero as well and, thus, $\sum_i S_i^V + R_i^K + R_i^V = 1$. %
Second, if there is no vaccination at all, i.e., $\nu_i \equiv 0$ for all~$i$, all vaccinated compartments stay empty and, therefore, $\sum_i S_i + R_i^K + R_i^U = 1$. %
The third case is a vaccination stop at some time~$t_2$, i.e., $\min_i \nu_i(t) > 0$ for $t \in [t_1, t_2)$ and $\max_i \nu_i(t) = 0$ for $t \geq t_2$. %
Here, one can infer $\sum_i S_i + S_i^V + R_i^K + R_i^U + R_i^V = 1$. %
Note that in a deterministic model like ours the compartments do not empty in finite time, e.g. $I_i(t) > 0$ for all $t \geq 0$ if $I_i(0) > 0$. %

Stability of an equilibrium can be characterized in terms of the largest real part of eigenvalues of the Jacobian $\nabla_x f(\bar{x}, \bar{u}, p)$. %
If the latter is negative, the equilibrium is stable, otherwise it is not. %
In order to study the stability numerically, we consider the equilibrium~$\bar{x}$ with $S_2 \in [0,1]$ and $R^U_2 = R^K_2 = 0.5 (1 - S_2)$ (and the remaining compartments being empty). %
Note that the~$R$ compartments do not affect the dynamics, i.e. the ratio $R^U_2 / R^K_2$ is irrelevant. %
Figure~\ref{fig:stability_of_equilibrium} (left) depicts the largest real part of eigenvalues of the Jacobian $\nabla_x f(\bar{x}, \bar{u}, p)$ with constant social distancing and without vaccination, i.e. $\bar{u} \equiv (\delta, \nu)$ with $\nu = 0$. %
If~$S_2$ is small, i.e. sufficiently many people have been exposed, the equilibrium is stable; otherwise adding infected people to the system might cause (another) outbreak. %
The vertical dotted black lines indicate herd immunity depending on the contact restrictions. %
Moreover, Figure~\ref{fig:stability_of_equilibrium} (right) shows that stability of the equilibrium depends on whether herd immunity has been achieved or not. %
Note that the real part of the leading eigenvalue of the Jacobian is strictly positive once the reproduction number~$\mathcal{R}_0^\mathrm{NGM}$ is larger than one. %

Moreover, once $S_2$ is sufficiently small (below approximately~$0.3$), the system has reached a robust invariant set, i.e. even without enforcing counter measures, there will not be another outbreak. %
The investigation of invariant regions and robust sets is outside the scope of this paper. %
For a set-based approach to maintain hard infection caps (during an outbreak of dengue fever) we refer to~\cite{EstALetal20}. %

\section{Minimizing the total number of fatalities}\label{app:hosp_min}
In this appendix, we present optimal vaccination policies based on minimizing the total number of people who have left intensive care units over a period of 2~years. In our model, this number is proportional to the total number of fatalities, see Section~\ref{sec:novel:seiphr:model}. We assume a constant level of social distancing, and we demonstrate that the qualitative nature of the optimal vaccination policies changes depending on the level of social distancing.

The optimal vaccination profiles are obtained by solving the optimal control problem
\begin{subequations}\label{nu_problem}
	\begin{align}
		\label{nu_problem:obj}
		\min_{\nu} \quad & H^C(t_f) + \kappa \left\| \nu \right\|_2^2 \\
		\label{nu_problem:hc:ode}
		\text{subject to} \quad & \dot H^C(t) = \sum_{i=1}^{n_g} \sigma (H_i(t) + H_i^V(t)), \quad H^C(0) = 0 \\
		\label{nu_problem:ode}
		& \dot{x}(t) = f(x(t), u(t), p), \quad x(0) = x^0 \\
		\label{nu_problem:vacc}
		& n_\mathrm{pop} \cdot \int_0^t{ \sum_{i=1}^{n_\mathrm{g}} \nu_i(s) V_i(s) \, \mathrm{d} s} \leq V^\mathrm{max} \cdot t \\
		\label{nu_problem:sd}
		& \delta(t) = \delta^c, \quad \forall \, t \geq 0 \\
		&\nu(t)    = \nu(k\Delta t),    \quad t\in[k\Delta t, (k+1)\Delta t[, \quad k = 0, \ldots, K-1\label{eq:nu_problem:zoh}
	\end{align}
\end{subequations}
Here,~$H^C(t)$ is the cumulative number of people who have left intensive care units up until time~$t$, and its temporal evolution is given by~\eqref{nu_problem:hc:ode}. The objective is to minimize this quantity at the end of a 2-year time period plus a regularization term, and we choose $\kappa = 5 \cdot 10^{-6}$ which is negligible compared to $H^C(t_f)$. Furthermore,~\eqref{nu_problem:ode}-\eqref{nu_problem:vacc} are the SEIPHR model~\eqref{eq:model-1}-\eqref{eq:model-2} and the constraint on the number of available vaccines, respectively. Finally, the level of social distancing is specified in~\eqref{nu_problem:sd}, and the vaccination profile is constrained to be constant over time intervals of one week by~\eqref{eq:nu_problem:zoh}.

\begin{figure}[htbp!]
	\centering
	\includegraphics[scale=0.4]{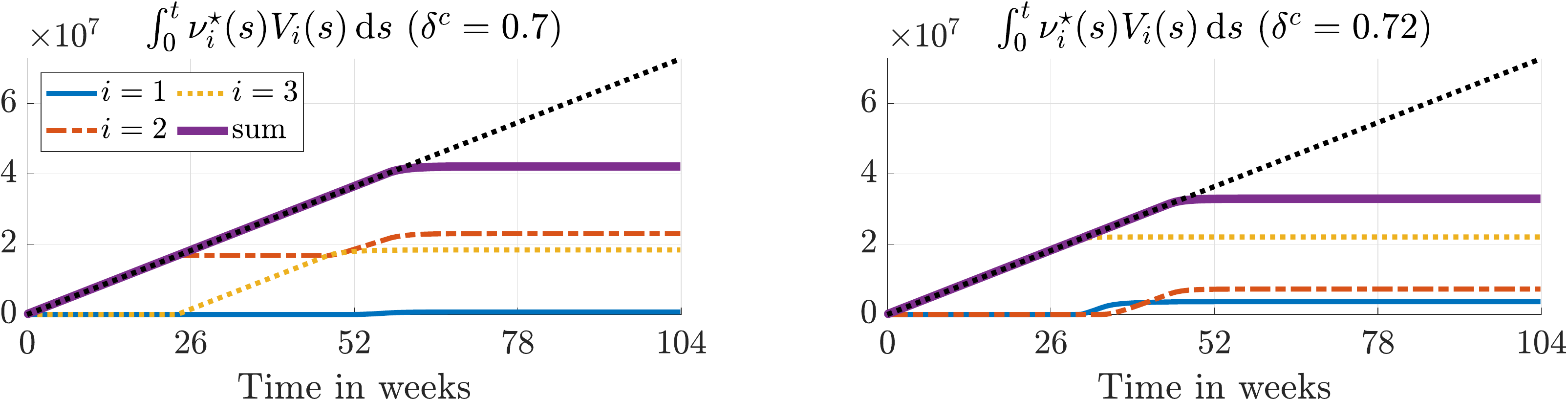}
	\caption{Optimal age-differentiated vaccination profiles for children ($i=1$), adults ($i=2$), and elderly ($i=3$) given different levels of social distancing. Here, we consider a vaccination success rate of $q = 0.9$ and a daily number of available vaccines of $V^\mathrm{max} = 10^5$.}
	\label{fig:nu:profile}
\end{figure}

\begin{figure}[htbp!]
	\centering
	\includegraphics[scale=0.4]{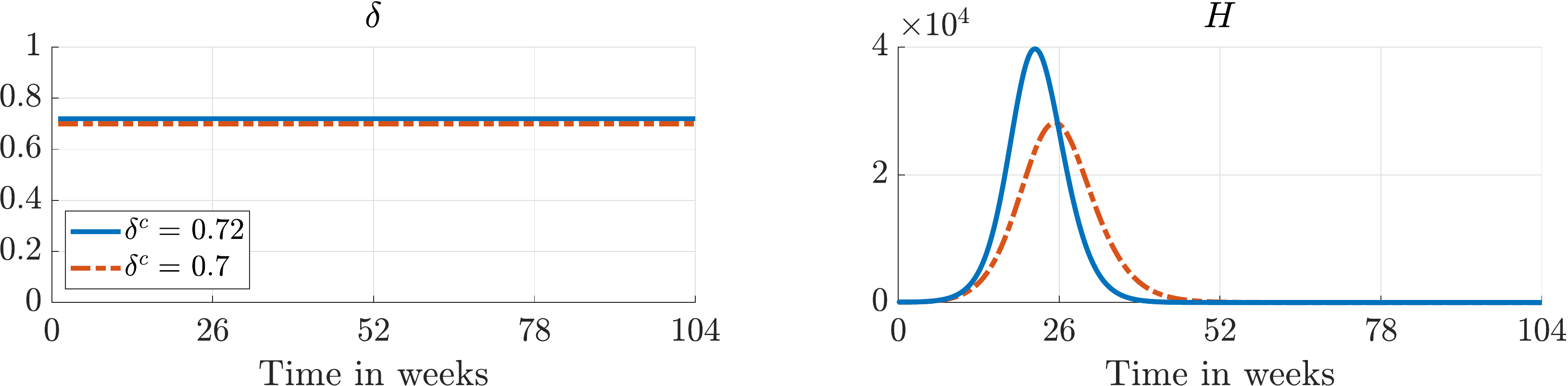}
	\caption{The amount of social distancing (left) and the number of hospitalized (right) corresponding to the optimal vaccination profiles shown in Figure~\ref{fig:nu:profile}.}
	\label{fig:nu:hospitalized}
\end{figure}

Figure~\ref{fig:nu:profile} shows the optimal vaccination profiles for $\delta^c = 0.70$ (left) and $\delta^c = 0.72$ (right). For $\delta^c = 0.70$, the adults are vaccinated before the elderly as was the case when minimizing the amount of social distancing in Section~\ref{sec:open:loop} and when sufficiently long prediction horizons were used in Section~\ref{sec:mpc}. In this case, roughly 280,000 were admitted to intensive care units (the objective function value is 280,956). For $\delta^c = 0.72$, the elderly are vaccinated before the adults as is the case when very short prediction horizons were used in Section~\ref{sec:mpc}. For this profile, around 300,000 went to intensive care units (and the objective function value is 300,238). Furthermore, Figure~\ref{fig:nu:hospitalized} shows the number of hospitalized. The peak value is lower when $\delta^c = 0.7$ and the adults are vaccinated first. This is due to the stricter social distancing measures and because adults, who have the highest contact rates, are vaccinated first.

If $\delta^c$ is above $0.72$, the profiles are qualitatively similar to that obtained with $\delta^c = 0.72$, and if $\delta^c$ is below $0.70$, the profiles are similar to that obtained with $\delta^c = 0.70$. Presumably, the two qualitatively different profiles become equally optimal for some value of $\delta^c$ between $0.70$ and $0.72$. However, because of the presence of at least two local minima, investigating this further is numerically challenging. Therefore, based on the qualitative nature of the optimal vaccination profiles in Figure~\ref{fig:nu:hospitalized}, we compare two heuristic profiles for either vaccinating the elderly or the adults first. In both cases, children are vaccinated last.
The heuristic vaccination profiles are based on the remaining number of vaccines at time $t_k$,
\begin{align}
	V_{\mathrm{rem}, k} &= \frac{V^{\max}}{n_\mathrm{pop}}t_{k+1} - \int_0^{t_k} \sum_i^{n_g} \nu_i(s) V_i(s) \mathrm ds.
\end{align}
The first term is the number of vaccines available at time $t_{k+1}$ and the second term is the total number of vaccines administered until time $t_k$. In the first profile, the elderly (age group~3) are vaccinated first. Furthermore, we consider piecewise constant vaccination rates which are constant over periods of one week as for the optimal vaccination profiles. Consequently,
\begin{subequations}\label{eq:vacc:elderly:first}
	\begin{align}
		\nu_3(t)
		&=
		\begin{cases}
			\frac{V_{\mathrm{rem}, k}}{V_3(t_k)(t_{k+1} - t_k)} & V_3(t_k) > \alpha N_3, \\[1mm]
			0 & \mathrm{otherwise},
		\end{cases} \\
		\nu_2(t)
		&=
		\begin{cases}
			\frac{V_{\mathrm{rem}, k}}{V_2(t_k)(t_{k+1} - t_k)} & V_3(t_k) \leq \alpha N_3 \text{ and } V_2(t_k) > \alpha N_2, \\[1mm]
			0 & \mathrm{otherwise},
		\end{cases} \\
		\nu_1(t)
		&=
		\begin{cases}
			\frac{V_{\mathrm{rem}, k}}{V_1(t_k)(t_{k+1} - t_k)} & V_3(t_k) \leq \alpha N_3 \text{ and } V_2(t_k) \leq \alpha N_2 \text{ and } V_1(t_k) > \alpha N_1, \\[1mm]
			0 & \mathrm{otherwise},
		\end{cases}
	\end{align}
\end{subequations}
for $t\in [t_k, t_{k+1}[$. The objective is to vaccinate $100(1 - \alpha)$ percent of each age group. As $V_i(t)$ is monotonically non-increasing, the above profile is guaranteed to satisfy the constraint~\eqref{nu_problem:vacc}. In the second profile, the adults (age group 2) are vaccinated first, i.e.,
\begin{subequations}\label{eq:vacc:adults:first}
	\begin{align}
		\nu_2(t)
		&=
		\begin{cases}
			\frac{V_{\mathrm{rem}, k}}{V_2(t_k)(t_{k+1} - t_k)} & V_2(t_k) > \alpha N_2, \\[1mm]
			0 & \mathrm{otherwise},
		\end{cases} \\
		\nu_3(t)
		&=
		\begin{cases}
			\frac{V_{\mathrm{rem}, k}}{V_3(t_k)(t_{k+1} - t_k)} & V_2(t_k) \leq \alpha N_2 \text{ and } V_3(t_k) > \alpha N_3, \\[1mm]
			0 & \mathrm{otherwise},
		\end{cases} \\
		\nu_1(t)
		&=
		\begin{cases}
			\frac{V_{\mathrm{rem}, k}}{V_1(t_k)(t_{k+1} - t_k)} & V_2(t_k) \leq \alpha N_2 \text{ and } V_3(t_k) \leq \alpha N_3 \text{ and } V_1(t_k) > \alpha N_1, \\[1mm]
			0 & \mathrm{otherwise},
		\end{cases}
	\end{align}
\end{subequations}
for $t\in [t_k, t_{k+1}[$.

\begin{figure}[htbp!]
	\centering
	\includegraphics[scale=0.4]{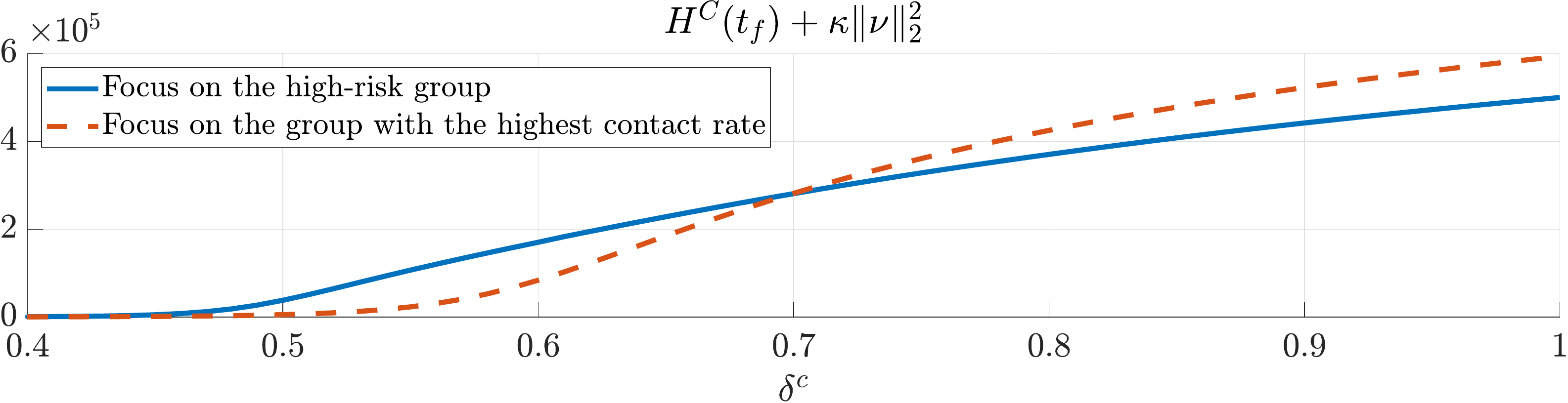}
	\caption{Values of the objective function in~\eqref{nu_problem:obj} obtained by vaccinating the elderly first using~\eqref{eq:vacc:elderly:first} and by vaccinating the adults first using~\eqref{eq:vacc:adults:first}. The vaccination success rate is $q = 0.9$ and the daily number of available vaccines is $V^\mathrm{max} = 10^5$.}
	\label{fig:nu:profile:heuristic}
\end{figure}

Figure~\ref{fig:nu:profile:heuristic} shows the value of the objective function in~\eqref{nu_problem:obj} obtained with the two heuristic vaccination profiles. %
For the optimal profiles, only a given percentage of each age group is vaccinated before starting to vaccinate the next. Therefore, we choose $\alpha = 0.1$. For the heuristic profiles, there is a cross-over point around $\delta^c = 0.7$. For values above, lower objective function values are obtained by vaccinating the elderly first and vice versa. When $\delta^c$ becomes sufficiently low, the two heuristic profiles result in the same objective function values. %
Figure~\ref{fig:nu:profile:heuristic} suggests that the objective function for the optimal strategy is not differentiable at the cross-over point.
We believe that it is optimal to vaccinate the elderly first because in this scenario the epidemic spreads too fast for vaccination of adults to prevent too many casualties. %
Although the objective function values corresponding to the optimal vaccination profiles shown in Figure~\ref{fig:nu:profile} are below the values obtained with the heuristic profiles, we expect that the conclusions carry over because of the similarity between the profiles.

	\bibliographystyle{unsrt}
	\bibliography{references}

\begin{thebibliography}{10}
\bibitem{Par20}
A.~Park.
\newblock Yes, we have {COVID}-19 vaccines that are 95\% effective. {B}ut that
  doesn't mean the end of the pandemic is near.
\newblock {\em Time}, 2020.
\end{thebibliography}
	
\end{document}